\documentclass[11pt,preprint]{aastex}

\newcommand{\simgt}{\hbox{\rlap{\raise 0.425ex\hbox{$>$}}\lower 0.65ex\hbox{$\sim$}}}
\begin{document}

%\shortauthors{NOLLENBERG ET AL.}
%\shorttitle{DUST IN GALAXY CLUSTERS}
%%Insert a gray "draft'' (with date) across page.
%\special{!userdict begin /bop-hook{gsave 340 50 translate
%20 rotate /Times-Roman findfont 30 scalefont setfont
%0 0 moveto 0.9 setgray (Draft \today ) 
%		       show grestore}def end}

\title{DETERMINATION OF REDDENING AND EXTINCTION DUE TO DUST IN $APM$
GALAXY CLUSTERS}
%\altaffilmark{1}}

\author{Joshua G. Nollenberg, Liliya L.R. Williams}
\affil{Astronomy Department, School of Physics and Astronomy,
	University of Minnesota, 116 Church St. S.E.,
	Minneapolis, MN 55455}
\email{manawa@astro.umn.edu, llrw@astro.umn.edu}

\author{Steve J. Maddox}
\affil{School of Physics and Astronomy, University of Nottingham, 
	Nottingham NG7 2RD, UK}
\email{Steve.Maddox@nottingham.ac.uk}

%\altaffiltext{1}{Based on observations with ISO, an ESA project with
%instruments funded by ESA member states (especially the PI countries:
%France, Germany, the Netherlands, and the United Kingdom) and with the
%participation of ISAS and NASA.}

%01234567890123456789012345678901234567890123456789012345678901234567890123456789

\begin{abstract}
Existing observations are consistent with rich clusters of galaxies
having no dust on $\;\simgt\;$Mpc scales, while galaxy groups most 
probably do have dust distributed over $\lesssim$ Mpc scales. Dust in
groups accounts for the observed redshift asymmetries of their galaxy
distributions, and about $E(B-V)\sim 0.1-0.2$ mag of reddening. Motivated by 
these results, we develop a new technique for 
determining the degree of reddening and extinction due to widely-distributed 
dust in nearby moderately rich and poor galaxy clusters. The method compares 
the color-magnitude plane distributions of galaxies from cluster and 
control regions on the sky, where control regions are assumed to be 
unaffected by 
dust. The method is statistical in nature; it can distinguish between 
uniformly, non-uniformly, and clumpily distributed dust, and can determine 
the amount of reddening and obscuration without a priori assuming an 
$A_R/E(B_{J}-R)$ ratio. We apply the method to nearby, $z\le 0.08$, medium 
rich and poor APM galaxy clusters. We detect no dust in these on $1.3$ Mpc 
scales (we assume $h = 0.75$), and derive 99\% confidence 
upper limits on extinction of 
$A_{R} = 0\fm 025 $ and reddening of $E(B_{J} - R) = 0\fm025~($which corresponds
to $E(B - V) \approx 0\fm 02)$.
We test the method using clusters whose galaxies have been artificially
reddened and obscured by various amounts, and conclude that it robustly 
recovers the input values for reddening, its distribution, and
the ratio of total-to-selective extinction. The method can be applied to any 
set of galaxy clusters or groups constructed out of homogeneous
and uniform two-color galaxy catalogs.
\end{abstract}

\keywords{galaxies:clusters: general - galaxies: fundamental parameters (color)
 - catalogs - techniques: photometric}

\section{Introduction}
\label{intro}

Dust inside a rich galaxy cluster's Intracluster Medium (ICM) would be subjected to 
harsh conditions.  Parameters of the hot, X-ray component of the ICM in galaxy
clusters have been measured by numerous studies. Typical temperatures 
are generally observed in the range $ T_{gas} \sim 10^{7} - 10^{8} K $, 
which is equivalent to thermal energies of $ kT \sim 2 - 14$ keV \citep{B99}.  
%though typically the gas density scales as
%$ \rho_{gas} \sim \left[ 1 + \left( \frac{r}{R_{c}} \right)^{2} 
%\right]^{-3\beta / 2}, $ where $\beta = \mu m_{P} \sigma_{DM}^{2} / k_{B} T \approx 0.8$
%\citep{NFW95}. 
Grains, which have typical molecular bonding 
potentials on the order of a few tenths to a few eV, are likely to 
dissociate by sputtering due to collisions with thermal electrons.  
The dust grain sputtering timescales are dependent on the size of the grains, 
$ a$ and the electron density at a given location in the ICM,
$ n_{e}(r) $:
\begin{equation} 
	\tau_{sp}(a, r) \approx 10^{6} \cdot \frac{a(\mu m)}{n_{e}(r)} yrs,
\end{equation} 
\noindent for graphite, silicate or iron grains \citep{DS79}, 
Electron densities on the order of $ n_{e} \sim 10^{-3} \cdot h^{1/2}  cm^{-3} $
have been measured \citep{JF92}, so typical
sputtering timescales are on the order of $\tau_{sp} \sim 10^{6} - 10^{9} yr.$
The shortest sputtering timescales correspond to grains located in the densest
regions of the ICM.
To estimate whether there will be any dust in rich clusters, 
dust destruction timescales have to be compared to dust injection timescales.
Dust could conceivably be introduced into the ICM
through several processes including ram pressure stripping of galaxies as they
travel through the ICM, the accretion of primordial dust, galaxy or cluster
mergers and collisions, blowout from galaxies that experience multiple 
supernov\ae ~ or intense starbursts, or cooling flows \citep{PTFV00}.  
Many of these processes have timescales on the order of
$ \tau \sim 10^{8} - 10^{9} yrs$, which is comparable to the longest
sputtering timescales.  Each of these modes of dust insertion affect different
spatial scales and locations within clusters.  For example, cooling flows
occur near the central regions of a cluster's gravitational potential while
accretion processes such as mergers affect the outer portions of the ICM.
Furthermore, hydrodynamical
processes may allow dust to reside in clumps, which would be self-shielded from 
ionizing radiation and electron collisions, resulting in a longer effective 
sputtering timescale and a smaller covering factor.  
Therefore, the amount and distribution of dust in a rich galaxy cluster may also
be dependent on the history of the various deposition processes in an 
individual cluster.

The nature of the extinction and reddening from dust in clusters 
may also depend heavily
on environment because the lifetimes of grains with some types of
chemistries and structures may be longer than the lifetimes of others under
the same ambient conditions.  In addition, deposition processes
could act as filters.  Grain acceleration by radiation pressure from
starlight, for example, could impose a bias of grain radius and mass on 
particles entering the ICM, from galaxies.  
In process, this could result in differences between the
initial and resulting extinction curves of the affect grain populations.
Certainly, the ratio of total-to-selective extinction, $R_{V} = 
A_{V} / E(B - V)$ is known to vary in the 
range $3 \lesssim R_{V} \lesssim 6$ within the Milky Way \citep{M90} and
more strongly in other galaxies, where $1.5 \lesssim R_{V} \lesssim 7.2$
\citep{f99}.  These facts together with the uncertainties in 
grain models for dust in the ICM require a search for dust in the ICM to be 
flexible in terms of detecting combinations of varying degrees of reddening 
and obscuration.  

\begin{figure}
%\vspace{18cm}
\vskip-3.0cm
\epsscale{0.9}
\plotone{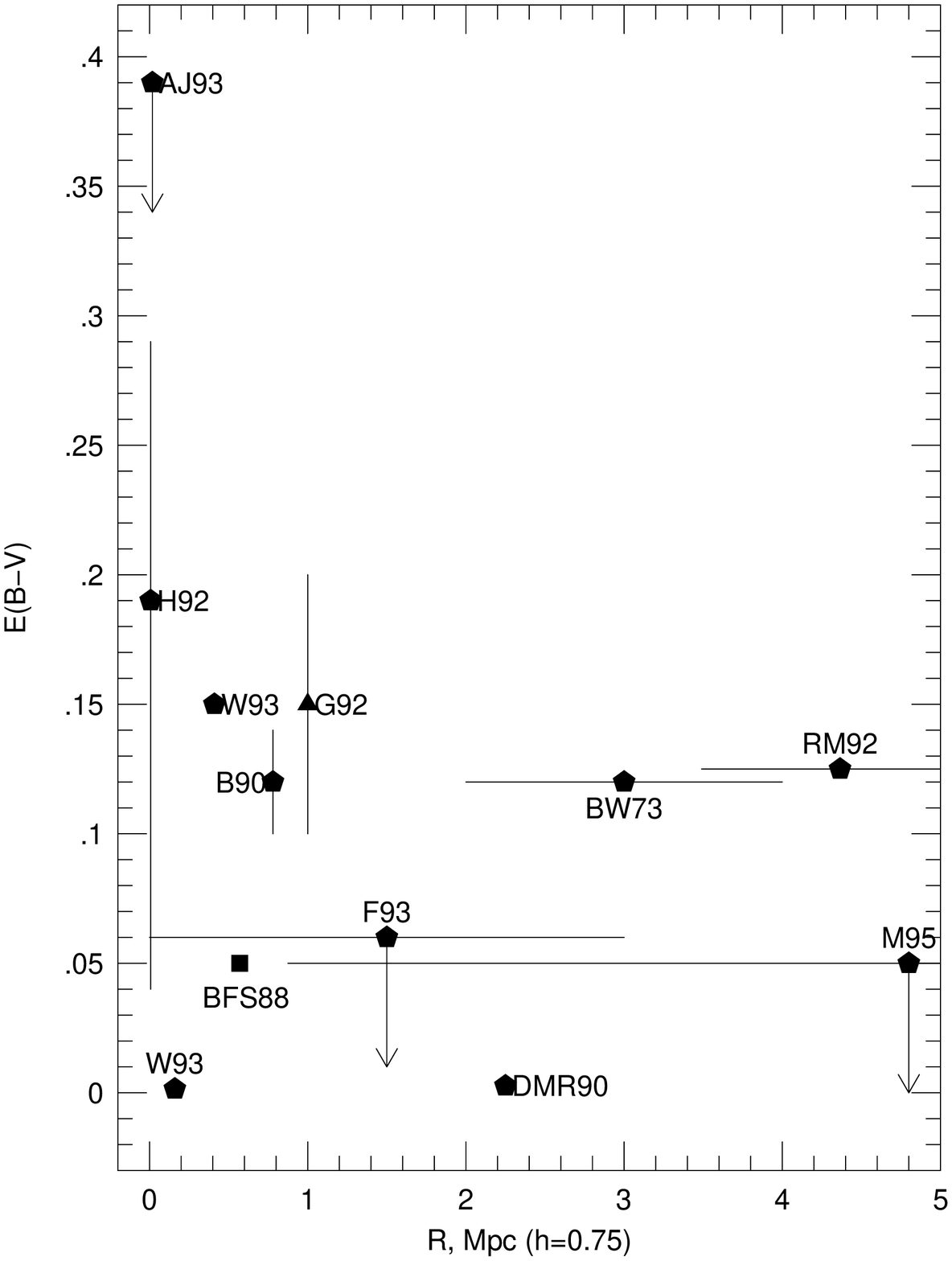}
%\special{psfile="abs6m025.ps" angle=-90 vscale=65 hscale=65 voffset= 50 hoffset=-30}
%\special{psfile = "plots/Fig1.ps" angle=0 vscale=75 hscale=75 voffset=0
%hoffset=0}
\vskip-1.8cm
\caption{Compilation of results from searches for dust in galaxy
clusters. Pentagons correspond to studies of rich clusters, squares represent
moderately rich and poor clusters, and the triangle represents galaxy groups. 
Horizontal ``error-bars'' indicate the spatial extent of dust measurements; 
arrows indicate upper limits. Some studies did not quote their results in
terms of color excess; in that case we used approximate conversion formulae,
$M_{dust}/M_\odot=2\times 10^{11} A_B\; (R/{\rm Mpc})^2$ \citep{BFS88}, and 
$A_B=4\,E_{B-V}$. There are two points labeled Wise et al. (1993): the
no-extinction point is for the majority of clusters in their study, while the
other point represent the findings for a small handful of clusters.}
\label{compil}
\epsscale{1.0}
\end{figure}

Figure~\ref{compil} gives a compilation of the results from prior 
searches for dust in rich and poor galaxy clusters, as well as groups. 
There is some observational evidence for the existence of dust in the central 
regions of clusters.
For example, using IRAS images of 56 clusters, 
\citet{WOBR93} found two clusters with Far-IR color excesses that are
probably due to dust. 
Meanwhile, \citet{H92} derived an average value of the differential reddening of 
$E(B - V) \sim 0\fm19$ on the basis of observed UV-to-optical emission line ratios
(Ly$\alpha / $H$\alpha$) in a sample of 10 Abell cooling flow clusters.
Furthermore, measurements of X-ray spectra have indicated surprisingly high 
metallicities, $Z \sim 0.2 - 0.5\,Z_{\sun}$, in clusters \citep{B99,AM98,VD95}.
Yet it should also be noted that \citet{AJ93} failed to detect any 
submillimeter dust emission from the central few tens of kpc 
in 11 cooling flow clusters.

Evidence for dust
in regions outside the central $\;\simgt\; 0.5$~Mpc is inconclusive.
Past searches for dust on $\simgt$ Mpc scales incorporated several techniques.  
\citet{Z62}, \citet{KL69}, \citet{dVC72}, and \citet{BW73} looked 
for the extinction of light from distant galaxy clusters by intervening 
foreground clusters.  
The observed depletion in the number density of background clusters led
\citet{Z62} to suggest that there 
may be as many as 0.4 magnitudes of extinction in the Coma Cluster. 
\citet{BW73} estimate that to explain the dearth of background clusters,
foreground clusters must produce about 0.12 magnitudes of reddening over
radii of few Mpc.
%\citet{dVC72} indicated that there may be as many as 0.3 magnitudes of extinction
%toward the center of the Local Galactic Supercluster. 
However, the results of these studies
could have been affected by human bias: Distant galaxy clusters are much harder to 
detect if they are superimposed on foreground clusters, so the apparent
depletion of distant clusters in the directions of nearby clusters could be due
to causes other than dust. 

\citet{RM92} looked at the apparent anti-correlation of high-redshift QSOs with
nearby Abell clusters. They measured a deficit of $\sim 25 \%$ in the quasar
number counts within 60\arcmin ~of rich clusters and noted that 
such a dearth of QSOs
can be explained by $A_B = 0\fm 15$ of extinction in these clusters.
However, a follow-up work by \citet{M95} reached a different
conclusion. He found no difference in the colors of QSOs found within $1\degr$ 
in projection from Abell clusters to those farther than $3\degr$ away and he
placed a $90 \% $ confidence upper limit on reddening of $E(B-V)  = 0\fm05 $.  
\citet{B90} found that blue Tully-Fisher distances to Virgo cluster spirals
correlate tightly with their color excess, while no such correlation was seen
in field spirals. They suggest that intracluster dust with reddening of $\sim 0.12$
magnitudes would make more distant galaxies within Virgo appear redder.
However, \citet{F93} analyzed fits to the correlations between $B - V$
colors and the reddening-insensitive Mg$_{2}$ index (the ratio of the 
Mg $b$ line flux to the adjacent continuum flux that straddles the line) 
for a sample of elliptical
galaxies and found no evidence for reddening in either the Virgo or
the Coma clusters.  He placed $90\% $ confidence upper limits of 
$E(B - V) = 0\fm06$ and $0\fm05$ for the Virgo and Coma clusters, respectively.
\citet{DRM90} were unable to detect any IR emission from dust
in the Coma Cluster, despite the earlier optical estimates by \citet{Z62}.
In fact, the upper limit placed on the $100 \micron$ emission was $\sim 400$
times lower than the theoretical estimates of \citet{VK84}, which were based
on the interactions between dust and the hot ICM.  

While the amount of dust in extended regions of rich clusters is consistent
with zero, there is strong evidence that dust exists in galaxy groups, on 
$\lesssim 1$~Mpc scales. The distributions of group galaxies 
in redshift are known to be skewed towards positive velocities
\citep{GMGM92}. These redshift asymmetries 
are most easily explained if galaxy groups are
assumed to be in the process of collapse, and the near-side
infalling galaxies (redshifted with respect to group center) are less
obscured, and  hence more numerous, than the far-side (blueshifted) galaxies.
The color excess vs. redshift distribution of group galaxies confirms this
scenario.
A very different type of observation could also suggest the presence of dust
in groups. 
\citet{BFS88} find that UV-excess selected sources (mostly QSOs) are
anti-correlated with poor galaxy clusters and groups. They estimate that
this could be explained by absorption of $A_{V} \approx 0\fm2 $ extending
over $~\sim 4-10^\prime$. These values are compatible with the amount and
the range of distribution of dust determined from redshift asymmetries by
\citet{GMGM92}. However, the \citet{BFS88} observations could well be the
result of weak gravitational lensing of faint QSOs with shallow number 
counts slope \citep{CS99,Metal02}.

Motivated by these observations we decided to search for dust in the
population of objects intermediate between rich clusters and groups,
i.e. moderately rich and poor clusters of galaxies. To that end we
develop a new method for determining the reddening and
extinction due to dust in galaxy clusters, and apply it to a set of
APM galaxy clusters.

Even a relatively small amount of extinction in galaxy clusters would have 
profound implications for a wide 
variety of fields.  Significant amounts of extant dust would obscure background
objects, and could skew statistics in studies that involve collections of
distant galaxies, quasars and supernov\ae. 
Observations of light curves of high-redshift Type Ia Supernov\ae , 
~for example, have shown 
 that these supernov\ae~are generally dimmer in comparison with nearer ones
 than one would expect if one were to assume an Open or Einstein-de Sitter
 Universe \citep{sn98,R98}.   This has been
 attributed to the existence of a cosmological constant, $ \Lambda $,
 although
 \citet{A99} has suggested that this could be the result of 
 extinction by intervening grey dust.  

Studies of weak lensing (anti-)correlations between foreground galaxies and
high redshift QSOs would also be affected.  \citet{T80} and \citet{C81} were the 
first to point out that magnification bias may create statistical
associations between foreground galaxies and high-redshift QSOs.  
The presence of dust in clusters would tend to decrease the number of background
QSOs in a flux limited sample, thus complicating the interpretation of 
results \citep{Metal02,RWH94}.

\section{Data}
\label{data}

\subsection{APM Catalogue}
\label{apm}

Our galaxy data originated from APM measurements of 30~ $5\fdg 8 \times 
5\fdg 8 $ UKST sky survey fields.  
We selected a region of the sky away from the plane of the Galaxy, covering 
 an area that spans  $ \alpha \approx 10\fh 0
 - 15\fh 0 $ and $ \delta \approx -7\fdg 9 - +2\fdg9, $ 
 which is roughly a range of~ $l \approx 237\degr - 361\degr$ and
 $b \approx 36\degr - 50\degr$ in Galactic coordinates. Each of 
 these UKST fields contain photometry that was measured digitally 
 by the Automated
 Plate Machine at the University of Cambridge \citep{Irwinpage}, 
 from the UKST $B_{J}$ and the UKST SES R sky surveys, which have limiting 
 magnitudes of 22.5 and 21 magnitudes, respectively, and are complete, roughly,
 to magnitudes of $B_{J} \approx 20.5 $ and $R  \approx 19.5$.  Vignetting
 reduces the plate density of galaxies.  Its effects become more pronounced 
with distance away from plate center, but become severe only beyond 
$2\fdg 7$ from the center. As a result, we selected only
 those galaxies that lie within $2\fdg 7$ of the centers of their 
 respective plates for our analysis.  Objects for our
 study must have been detected and identified as `galaxies'
 in both the $B_J$ and $R$ bands.

 \subsection{APM Clusters}
 
The APM Galaxy Cluster Catalogue consists of clusters selected on
the basis of spatial overdensities in the APM Galaxy Catalogue
\citep{DMSE97}.  Centers of
galaxy clusters were found using a percolation technique in which all galaxy
pairs with angular separations on the fields less than 0.7 times the mean 
separation
were linked and assigned to the same group.  The centroid of galaxy groups
containing more than 20 galaxies was considered a candidate cluster center. 
A starting value for the estimated redshift of each cluster was assumed,
which gave estimates of various ordinal galaxy magnitudes
(e.g. the $1^{st}$ or the $3^{rd}$ 
brightest galaxies in the cluster).  These were adjusted
along with cluster richness, redshift, and centroid in an 
iterative routine, until convergence was achieved.
The result is a
catalogue of galaxy clusters in which each cluster was selected on the bases of
an impartial empirical algorithm, rather than by human hand.  
   
The APM clusters have a correlation length of approximately $r_{o} = 
14.3 h^{-1}$ Mpc, which is $\lesssim 0.5 $ times the 
correlation length of Abell Clusters \citep{D94,DMSE97}.  This implies
a spatial density of APM clusters $\sim 10$ times greater than that of Abell 
clusters, hence APM clusters are moderately rich to poor.
We have selected for our sample those clusters that lie on the UKST fields
mentioned in Section~\ref{apm} and have estimated redshifts $ z \leq 0.08. $
Since all APM clusters have the same assumed physical radius, $1.3$ Mpc
(assuming $h=0.75$),
their angular size on the sky varies from about $15\arcmin$ to $90\arcmin$.
Clusters which lay sufficiently far from the centers of their APM plates to 
extend, on the basis of their angular size, 
beyond the $2\fdg 7$ radius described in the last section were not
included in our sample.

\section{Method}
\label{proc}

If there is extant dust in nearby galaxy clusters, then more distant galaxies
along the same line of sight as the foreground cluster should be reddened and
obscured to a degree depending on the characteristics of the intervening dust.  
Because of issues such as grey dust, for example, mentioned in the 
Introduction,
we have developed a procedure to search for dust in galaxy clusters that is 
sensitive to any ratio of general to selective extinction, $R_{\lambda}$.  

Galaxy clusters were selected from the APM Catalogue with redshifts of 
$z \leq 0.08 $ in order to ensure that the clusters subtended an angle of the
sky large enough to contain a significant number of background dust-affected
galaxies, and a small number of foreground galaxies, at $z<z_{clust}$. The fraction 
of foreground galaxies is $\lesssim 10\%$, given our faint magnitude cutoff
and empirical estimates of redshift distribution of flux-limited samples of
APM galaxies \citep{MES96}. Selecting nearby clusters also ensures that
lensing would not be an issue, considering the fact that the background galaxies
are relatively nearby as well. 
For each APM Cluster two apertures were selected, the  ``Cluster Group''
and the ``Control Group''; 
the method determines the obscuration and reddening
of the Cluster Group galaxies {\it relative} to those of the galaxies in the 
Control Group. The Cluster Group was either a circular or annular aperture centered 
on a galaxy cluster's center (as indicated by the APM Cluster Catalogue) and 
containing all galaxies that lie within an angular radius equivalent to the
physical radius of $1.3$ Mpc at the distance of the cluster.
The Control Group contained all of the galaxies located inside a circular
annulus concentric with the Cluster Group, with an inner radius of $1.3$ Mpc
and an outer radius which ensured that the Cluster and Control Groups had
the same sky area. 
If the Cluster Group of a given cluster extended beyond $2\fdg 7 $ from plate 
center, the cluster was not used. If a portion of the Control Group's annulus 
fell beyond that radius, then that part of the annulus was removed, 
and the annular thickness of the remaining portion
was increased to ensure equal areas. 
The Cluster and Control Groups for each individual
cluster were confined to the same APM plate to avoid issues relating to
plate-to-plate magnitude calibration. Since individual clusters do not have
enough galaxies to be analyzed using our procedure, galaxies from many clusters 
were combined to yield merged Cluster and Control Groups 
(see Section~\ref{analysis}).

It should be noted that despite the equal areas of the Cluster and Control
Groups for the complete combined dataset, the Cluster Groups contained 
 more galaxies $(N_{clust} = 98059, $~or roughly 715 per cluster on average$)$ 
 than the Control Groups $(N_{cont} = 85884, $~or 627 per cluster on average$)$, 
 because in addition to 
foreground and background galaxies, the Cluster Groups also contained cluster 
galaxies. Our method required that the total number of galaxies in the Cluster 
and Control Groups be about the same. To that end, we tried masking out the 
central regions of Cluster Groups (which typically have the highest density 
of galaxies), and adjusting the size of the Control Groups accordingly. 
However, reducing the fractional difference between the number of galaxies 
in the Cluster and Control Groups to $\lesssim 1 - 3 \%$ (from $\approx 10\%$) 
had little effect on our determinations of  extinction and reddening.

\begin{figure}
\plotone{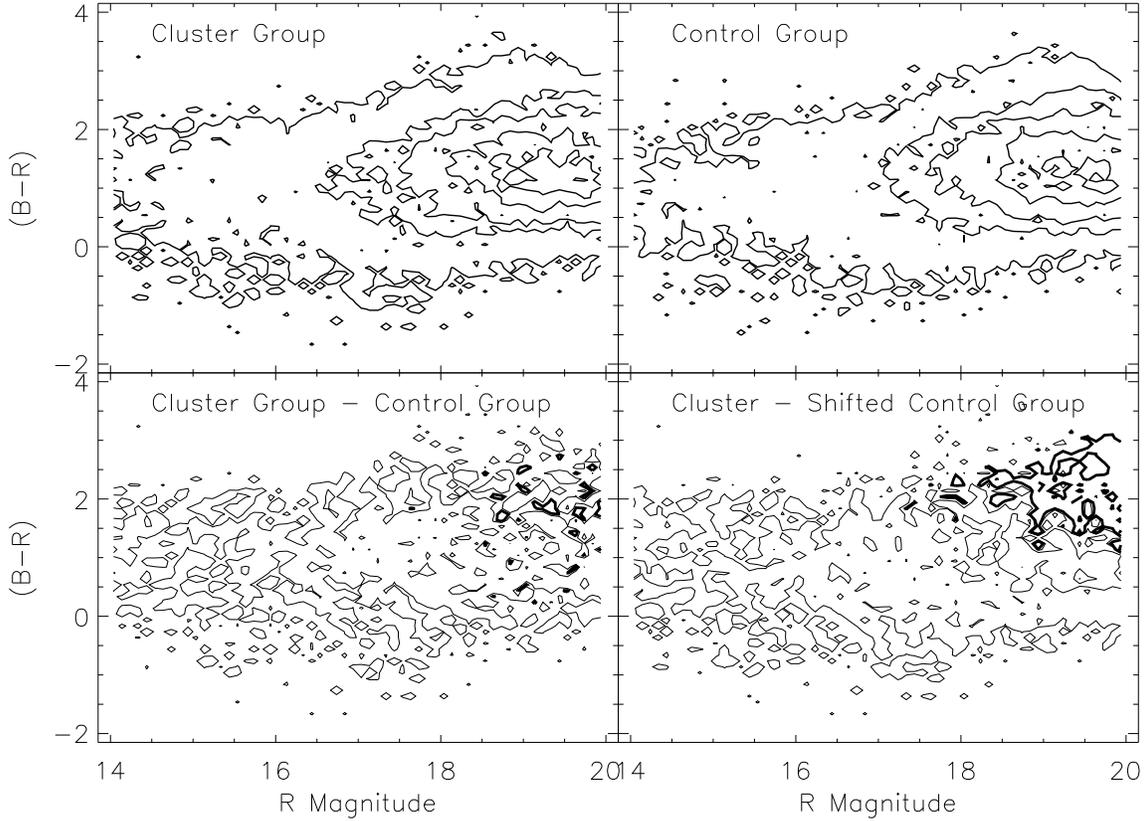}
\caption{{\it Top:} Distribution of Cluster and Control Group galaxies on the 
CM plane for the full sample.  Note that the former have more galaxies, 
as they should.  The background galaxies, i.e. those at $z > z_{clust}$ dominate
the overall galaxy count in each case.  Contour levels are 1, 2, 4, 8 and 12
galaxies per pixel.  {\it Bottom:} Examples of $n_{clust} - 
n_{cont}$ distributions used in computing eq. 2.  Thin  lines represent
positive contours in the density of galaxies in the CM plane, while thick lines
are negative contours.  The contour levels for the bottom panels are -4,
-1, 1, and 3 galaxies per pixel.  The Control Group in the lower right panel 
was shifted by a larger amount, $0\fm2$ in $R$ 
and $(B - R)$, for this figure than it was in our analysis for visualization.  
Note the oversubtracted region in the shifted case.}
\label{ndist}
\end{figure}

Next, we divided the distribution of galaxies in the color-magnitude (CM) plane 
of  both the Cluster and the Control Groups into pixels of size 
$\Delta_p R$, and $\Delta_p (B_J-R)$ (see top panels of Fig.~\ref{ndist}).
We present results for
$\Delta_p R=\Delta_p (B_J-R)=0\fm025$. Given the number density of APM galaxies,
this is the smallest reasonable pixel size.
The CM-plane pixel size determines the sensitivity, or `resolution' of our method. 
Let $n_{clust}(i,j)$ and $n_{cont}(i,j)$
%%%$n_{clust}(R, [B_{J}-R])$ and $n_{cont}(R, [B_{J}-R])$ 
be the observed pixellated CM-plane distributions of galaxies from the
Cluster Group and the Control Group, respectively.  For the combined analysis,
$n_{clust}$ and $n_{cont}$ have maximum values of 44 and 37 galaxies in one pixel, 
respectively.
We have selected rectangular subsets of the distributions,
limited on all four sides 
by the color cut $-2 \leq (B_{J} - R) \leq 4$ and magnitude cut of $14 \leq R
\leq 20$. The faint end of the
magnitude cut ensures a photometrically complete sample.
Dust in galaxy clusters would obscure
and redden the light coming from galaxies directly behind them, so that
background galaxies in the Cluster Groups would be displaced in the CM plane
by the corresponding reddening vector, while galaxies in the Control Group
would not. (Since the number of cluster member galaxies and foreground galaxies 
is small compared to the background galaxies, we assume that all galaxies
in the Cluster Group are background.)  Therefore
it is possible to search for dust by comparing the distributions
of galaxies on the CM plane from the Cluster and Control Groups.  

In the simplest case of constant reddening vector, the reddening and
extinction can be found by sliding the Control Group CM plane
distribution with respect to the that of the Cluster Group until 
the shapes of the two distributions match. 
The actual values of reddening and extinction would then
correspond to the pixel displacements $(k,l)$ that minimize
\begin{equation} \xi_{k,l} = \frac{1}{N(k,l)}\sum_{i,j}\frac{\mid
{n}_{clust}(i,j) - {n}_{cont}(i+k,j+l) \mid}
{{n}_{clust}(i,j) + {n}_{cont}(i+k,j+l)},\label{statistic}  \end{equation}
\noindent
where cell combinations in which $n_{clust}(i,j) + n_{cont}(i+k,j+l) = 0$ 
are not included in the analysis or the value of $N(k,l)$.
The bottom panels of Figure~\ref{ndist} shows two examples of 
$(n_{clus} - n_{cont})$ distributions, with specific values of color 
and magnitude shifts. Note that by applying shifts to the
Control Groups, we create regions on all four edges of the CM plane 
in which pixels from both arrays do not overlap. To compensate for
this effect the sum in eq.~\ref{statistic} is over the subset of the CM 
plane where the ${n}_{clust}$ and ${n}_{cont}$ distributions do intersect, 
and so $N$ is a function of the applied shift, $(k,l)$.
The greater the relative shift between the Cluster and Control Groups, 
the smaller the overlap area will be. The maximum shifts we apply in $R$ and 
$(B_J-R)$ are $\pm 0.1$ mag, which is a realistic upper limit on the amount 
of dust expected to be found in clusters (see Fig. 1). 
For a maximum shift $\pm 0.1$ mag in both CM-plane directions, 4700 pixels 
out of a total 57600 had to be removed; the fractional decrease in pixel 
number is small, $\sim 8\%$.

If the minimum of $\xi_{k,l}$ corresponds to pixel shifts of $(k_*, l_*)$,
then the reddening and extinction in the cluster are 
$A_R=k_*\,\Delta_p R$, and
$E(B_{J} - R) = \,l_*\, \Delta_p (B_{J}-R)$ respectively, while
the ratio of total-to-selective extinction is
$R_{R} = {A_{R}}/{E(B_{J} - R)}.$

If the distribution of dust in clusters is patchy, so that there is
a range of reddening and extinction values, then the procedure 
described above will recover the distribution of the $A_R$ and $E(B_{J} - R)$
values (averaged over the clusters), to within the resolution limit set
by the CM pixel size. However, if the dust is perfectly grey, and the
galaxy number counts follow a single power law, then the Cluster Group and
the Control Group CM distributions will look very similar except for 
normalization. In that case our method will not be able to detect the 
presence of dust. Regardless, this method
allows the simultaneous detection of multiple dust populations that have
different extinction curves, something that is not possible when
simply determining the average galaxy color in the sample.
In Section~\ref{tests} we demonstrate this using 
clusters which have been `artificially' endowed with dust. 

Each set of galaxy clusters that we examine using our procedure
contains at least 25 clusters, with most samples containing more than 100, 
which are widely distributed over $\sim 1000$ sq. deg.
of high Galactic Latitude. Assuming that the distribution of cirrus dust in 
our own Galaxy does not mimic the distribution of APM clusters, 
the effects of reddening and obscuration due to Galactic dust on galaxies 
in our Cluster and Control Groups should average out.

\section{Analysis of APM Clusters}
\label{analysis}

We generated lists of galaxies lying in the Cluster and Control
Groups for approximately 140 APM clusters.
The algorithm detailed in the previous section was used on three separate sets
of clusters: (i) the complete sample, (ii) sample with clusters whose apparent 
radii are $\le 30\arcmin$ (i.e. $z \geq 0.04$), and (iii) clusters with 
$r_{clust} > 30\arcmin$ ($z < 0.04$). Individual clusters cannot be analyzed
using our method due to the small number of galaxies per pixel in the CM plane.
In the case of the combined set, the average number of galaxies per pixel is on
the order of 10s, whereas the number per pixel for an average cluster would be 
on the order of 1 at most.

\begin{figure}
%\vspace{18cm}
\plotone{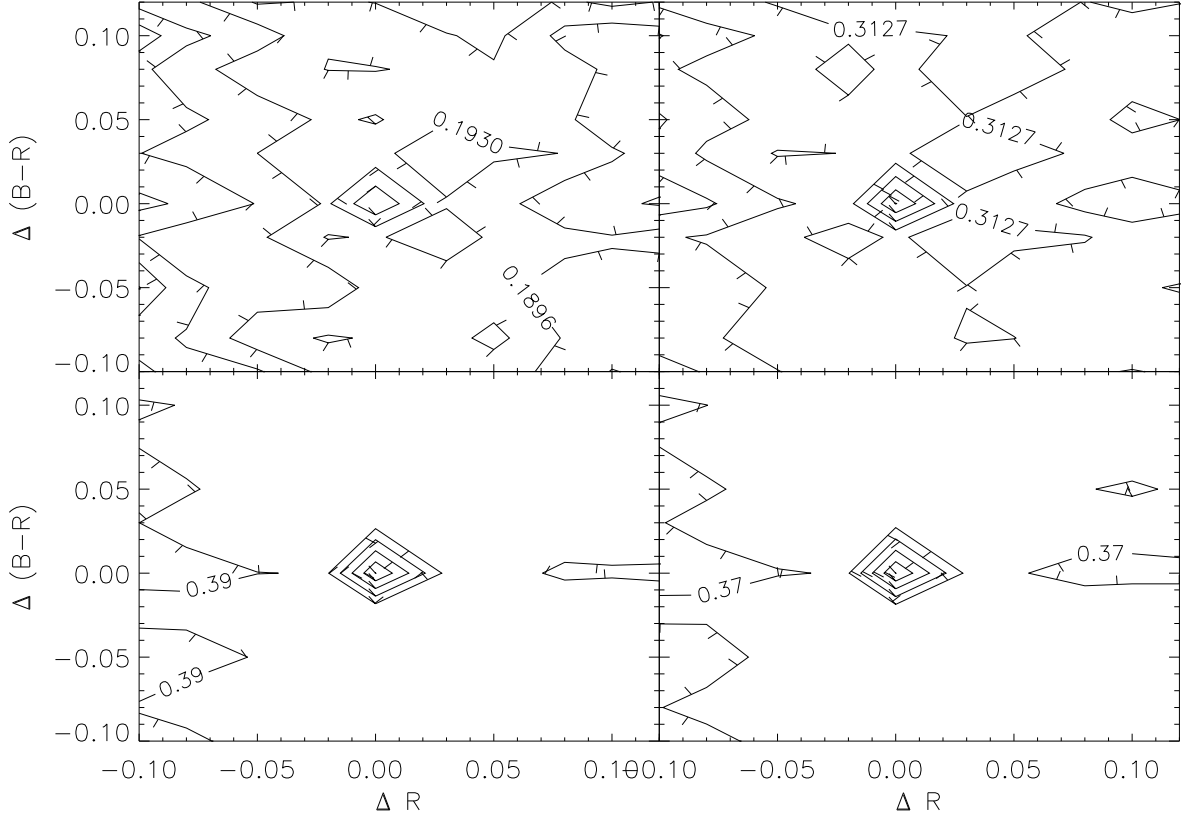}
%\special{psfile="abs6m025.ps" angle=-90 vscale=65 hscale=65 voffset= 50 hoffset=-30}
%\special{psfile = "plots/Fig2.ps" angle=0 vscale=75 hscale=75 voffset=0
%hoffset=0}
\caption{Contour plot of $\xi_{k,l}$ versus shifts in color and
magnitude for the combined sample of all galaxy clusters, plotted as
$\Delta R=k\,\Delta_p R$, and $\Delta E(B_{J} - R) = l\, \Delta_p (B_{J}-R)$.
The Inner and Outer Cluster Radii ($R_i$ and $R_o,$ respectively)
are given in units of Mpc ($h = 0.75$), along with the 
total number of Cluster and Control galaxies in each sample ($N_{clust}$ 
and $N_{cont}$,
respectively). The minimum value of $\xi_{k,l}$
for each of these plots corresponds to $\Delta(B-R) = \Delta R = 0$.
{\it Upper Left:} $R_i = 0, R_o = 0.33, N_{clust} = 8026, N_{cont} = 7901.$ 
{\it Upper Right: } $R_i = 0, R_o = 0.66, N_{clust} = 29003, N_{cont} = 28679.$ 
{\it Lower Left: } $R_i = 0, R_o = 1.3, N_{clust} = 98059, N_{cont} = 85884$. 
{\it Lower Right:} $R_i = 0.66, R_o = 1.3, N_{clust} = 69056, N_{cont} = 68666$.}
\label{trialsa}
\epsscale{1.0}
\end{figure}

\begin{figure}
%\vspace{18cm}
\plotone{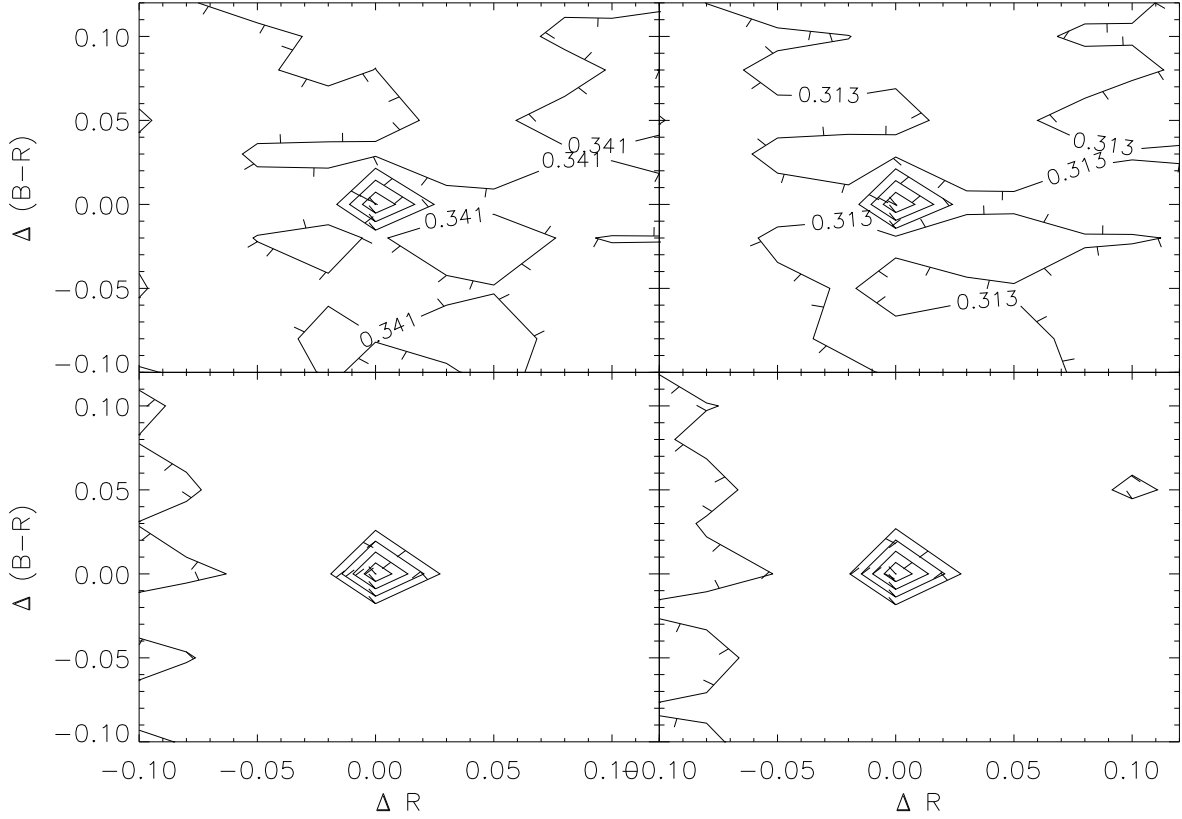}
%\special{psfile="abs6m025.ps" angle=-90 vscale=65 hscale=65 voffset= 50 hoffset=-30}
%\special{psfile = "plots/Fig2.ps" angle=0 vscale=75 hscale=75 voffset=0
%hoffset=0}
\caption{Similar to Fig.~\ref{trialsa}, however now galaxy clusters with apparent 
radii of $ \leq 30\arcmin$ (111 clusters) in our sample
are separated from those clusters with angular radii of $> 30\arcmin$ 
(26 clusters). 
Again, the minimum values of $\xi_{k,l}$ for each case correspond to 
$\Delta(B_J-R)=\Delta R=0.$
{\it Upper Left:} $r_{clust} \leq 30\arcmin, R_i = 0, R_o = 1.3,
	N_{clust} = 40708, N_{cont} = 33560$. 
{\it Upper Right: } $r_{clust} \leq 30\arcmin, R_i = 0.66, R_o = 1.3,
	 N_{clust} = 27739, N_{cont} = 27956$. 
{\it Lower Left: } $r_{clust} > 30\arcmin, R_i = 0, R_o = 1.3,
	N_{clust} = 57506, N_{cont} = 52642$. 
{\it Lower Right:} $r_{clust} > 30\arcmin, R_i = 0.66, R_o = 1.3,
	N_{clust} = 41300, N_{cont} = 40700$.}
\label{trialsb}
\epsscale{1.0}
\end{figure}

Each of the three sets of clusters were divided even further in order 
to search for dust on various size scales by removing the central, or the 
outer regions of the Cluster Groups. We show the results of our analysis 
in Figures~\ref{trialsa} and \ref{trialsb}, where we plot contour
levels of $\xi_{k,l}$ (eq.~\ref{statistic}) as a function of the applied shift in 
the CM plane, $\Delta R=k~\Delta_p R$, and $\Delta (B_J-R)=l~\Delta_p (B_J-R)$.
Figure~\ref{trialsa} shows the 
results from selecting various physical scales from our combined data set of 
roughly 140 APM clusters while Figure~\ref{trialsb} shows various subsets of 
clusters with radii $ \le 30\arcmin$ (upper panels) and $> 30\arcmin$ 
(lower panels).

In all our analysis of APM clusters, the minima in $\xi_{k,l}$ are always at 
$(0,0)$; in other words, we detect no extinction or reddening due to dust in 
any of the cases. Note that
the lowest contour level in the plots is not at 0, but typically at
$\sim 0.2-0.3$, consistent with what one should expect given the definition
of $\xi_{k,l}$ (eq.~\ref{statistic}) and the level of Poisson noise in the 
number of galaxies per CM-plane pixel.

To estimate the robustness of our results we implemented a bootstrapping 
algorithm. We randomly selected 100 sets of galaxy clusters from our sample 
(keeping the total number of clusters in each set the same, and equal to the 
original number), performed our analysis of each set separately, and tabulated 
the $\Delta R$ and $\Delta (B_J-R)$ values corresponding to the minima of
$\xi_{k,l}$. 
By selecting random combinations of clusters we are in effect augmenting 
some and suppressing other random subsets of the data.  The degree
to which the various subsets of the data agree serves to test the
robustness of our results.  We interpret the histogram of the number of
times that each pixel on the $(k,l)$ plane was found to correspond 
to $min(\xi_{k,l})$ over 100 permutations of our data
as the $\chi^{2}$ confidence limits for our sample of 
galaxy clusters.  The bootstrapping results indicate non-detection of 
widely-distributed dust in each of the trials shown in Figures~\ref{trialsa}
and \ref{trialsb}, with limits of $|\Delta R| < 0.025$ and 
$|\Delta(B_{J} - R)| < 0.025$ at a confidence level of $99 \% $, except 
for the noisy case in the Upper Left panel of Figure~\ref{trialsa}, 
which corresponds to no extinction or reddening ($|\Delta R| < 0.025$ and 
$|\Delta(B_{J} - R)| < 0.025$) at the $66\% $ level.

We note that in order to determine the average reddening of the Cluster 
Group compared to the Control Group galaxies, regardless of the ratio of 
total-to-selective extinction, it would have been sufficient to compute 
the difference in the average $\Delta (B_J-R)$ color of the two Groups,
instead of using eq.~\ref{statistic} statistic. The former, which uses
galaxy counts integrated along the magnitude direction, has the 
advantage of yielding results of higher statistical significance. The
disadvantages are twofold: First, the ratio of total-to-selective 
extinction, $R_R$ cannot be determined, and second, the properties of
non-uniformly distributed dust, of the type considered in the next 
Section, cannot be easily recovered.

\section{Confidence Tests and Sensitivity}
\label{tests}

In Section~\ref{analysis} we found that the $1.3$ Mpc regions around 
APM clusters do not appear to have any dust in them, compared to their 
surrounding Control annuli. Here we examine whether the technique we have 
developed is sensitive enough to detect reddening and obscuration if they were 
non-zero. We artificially introduce `dust' into the Cluster Group galaxies,
and leave Control Group galaxies unaltered. Several types of dust were used:
(i) uniformly distributed dust of constant total-to-selective extinction ratio, 
$R_R$; (ii) non-uniformly distributed dust of constant $R_R$ ratio;
(iii) patchily distributed dust, with distinct dust clumps, and unobscured areas.
Unless otherwise specified, we used the \citet{RL85} Interstellar Extinction Law
for all types of dust, 
and their ratios for $E(R - V)/E(B - V)$ and $A_R/A_V$ to calculate 
$R_R = 1.3$ given $R_V \approx 3.1$. Other values of total-to-selective 
extinction ratio could have been used. Our technique's ability to determine 
$R_R$ is independent of its value.

\begin{figure}
%\vspace{18cm}
\plotone{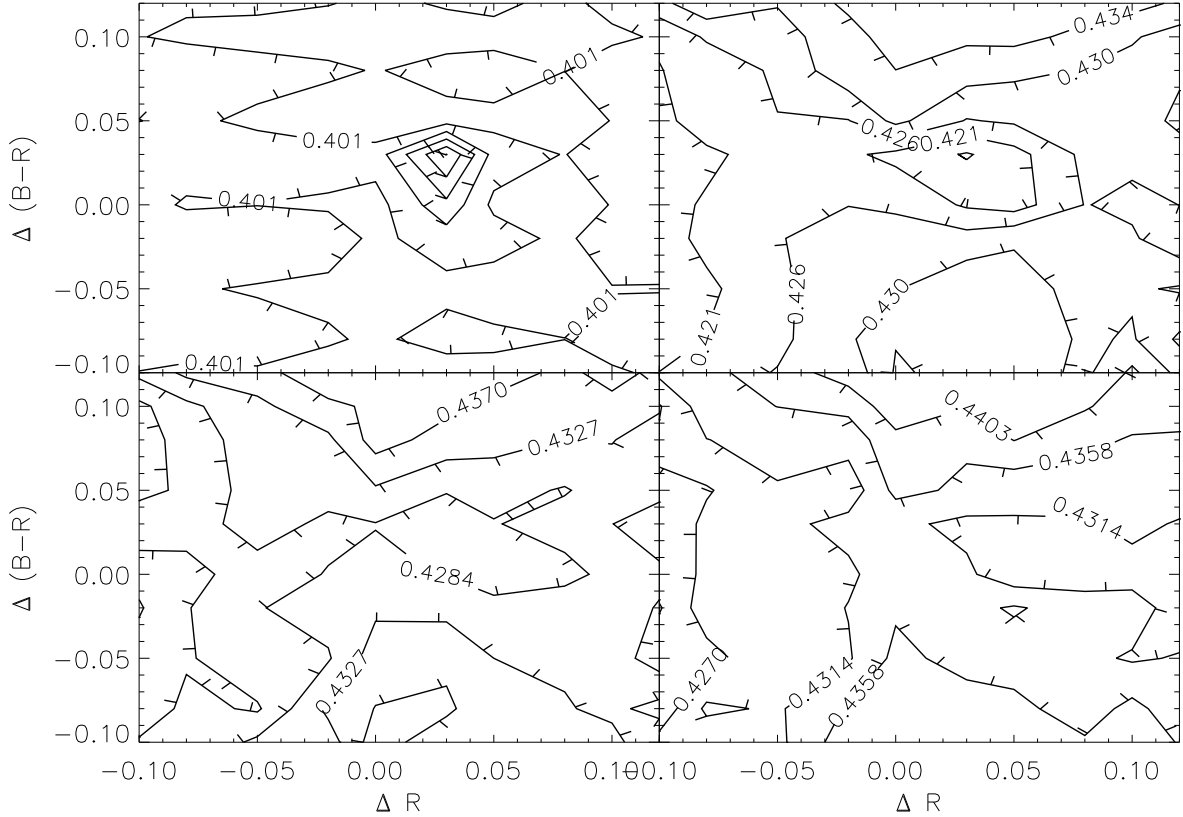}
%\special{psfile="abs6m025.ps" angle=-90 vscale=65 hscale=65 voffset= 50 hoffset=-30}
%\special{psfile = "plots/Fig2.ps" angle=0 vscale=75 hscale=75 voffset=0
%hoffset=0}
\caption{Testing our method: uniformly distributed `dust' was introduced into the 
Cluster Groups by artificially shifting the Cluster Group galaxies by 
$\Delta_a R = 0\fm025$ and $\Delta_a (B_J-R)=0\fm020$ (i.e. $R_R = 1.3$).
{\it Upper Left:} All
Cluster Group galaxies shifted by the same $\Delta_{a}R = 0\fm025$, to mimic
uniform distribution of dust. 
{\it Upper Right:} Cluster Group Galaxies were shifted in $R$ by an amount
randomly picked from a Gaussian distribution with $\langle \Delta_a R\rangle = 0\fm025$ 
with $\sigma_{a,R} = 0\fm025.$ 
{\it Lower Left:} Similar to Upper Right, but with $\sigma_{a,R} = 0\fm035.$  
{\it Lower Right:} Similar to Upper Right, but with $\sigma_{a,R} = 0\fm05$.}
\label{dist}
\epsscale{1.0}
\end{figure}

Since uniformly distributed dust would affect all background galaxies,
the first type of dust was implemented by shifting all the galaxies in
the Cluster Group by the same amount in magnitude and color.  
We chose $\Delta_a R=0\fm025$, and the corresponding 
$\Delta_a (B_J-R)=0\fm019$, where subscript $a$ stands for `artificial' dust.
This example of dust type (i) is shown in the Upper Left panel of Figure~\ref{dist}.
Within our resolution, the minimized statistic recovers the correct values of the shifts
in reddening and extinction.

Non-uniformly distributed dust would affect background galaxies to various 
degrees, so for the second type of dust we assumed an average 
value for extinction $\langle\Delta_a R\rangle$, as well as a dispersion 
in extinction, $\sigma_{a,R}$. The shape of the $\Delta_a R$ distribution 
that we applied was Gaussian. Individual galaxies from the Cluster Groups 
were assigned values of $\Delta_a R$ randomly from the distribution.
The corresponding color excess was determined using a fixed ratio or
$R_R = 1.3$ .
We present results for case (ii) with $\langle\Delta_a R\rangle = 0\fm025$, and
$\sigma_{a,R}=0\fm025, 0\fm035,$ and $0\fm05$ in the Upper Right and Lower 
panels of Figure~\ref{dist}. As before, the corresponding color shift for 
each galaxy was obtained assuming constant $R_R$ ratio.  From these cases, it is
clear that the wider the distribution of shifts, the more difficult it is
to recover the correct values of the extinction and reddening shifts.  This is 
due to the fact that the signal eventually becomes lost in the noise when
implementing wider and wider distributions.  Hence, it is necessary to use a
deeper catalog with a larger coverage area
in order to determine the distribution of extinctions and
reddenings inside a set of clusters where dust is distributed non-uniformly.

\begin{figure}
%\vspace{18cm}
\plotone{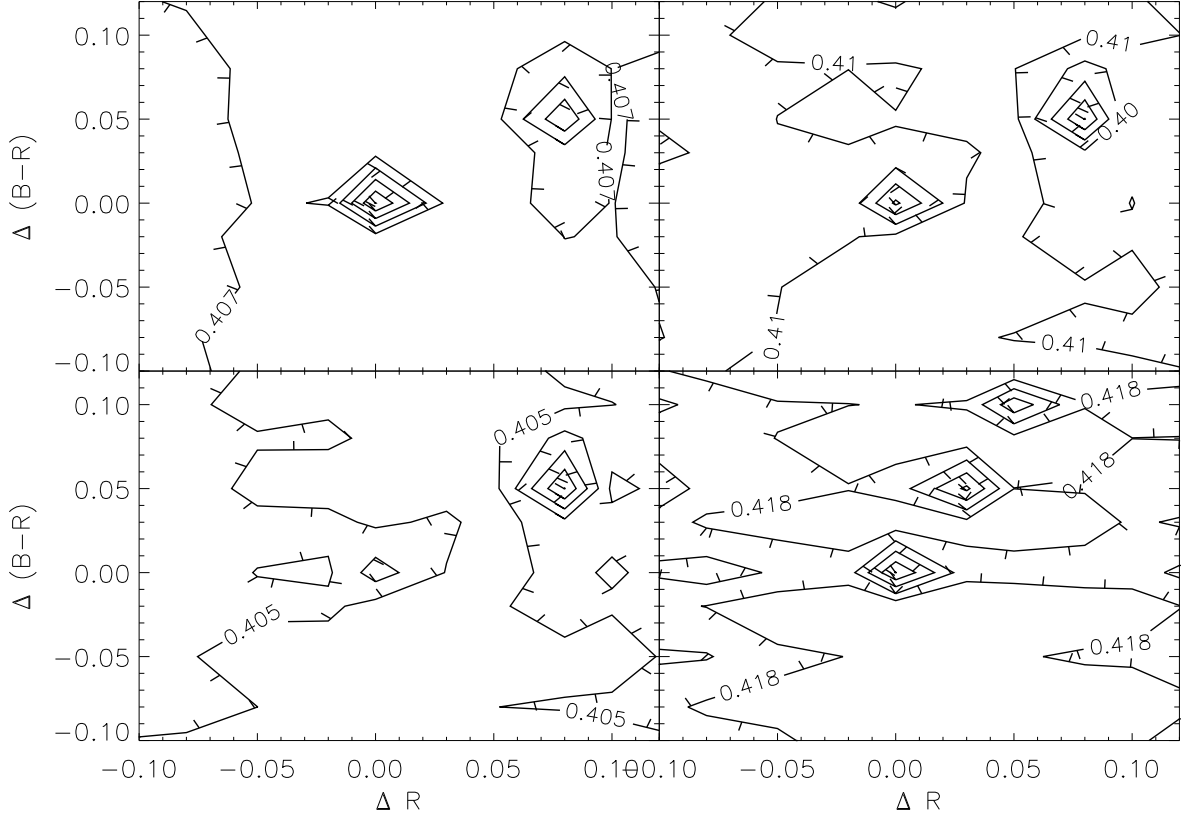}
%\special{psfile="abs6m025.ps" angle=-90 vscale=65 hscale=65 voffset= 50 hoffset=-30}
%\special{psfile = "plots/Fig2.ps" angle=0 vscale=75 hscale=75 voffset=0
%hoffset=0}
\caption{Tests of bi- and tri-modal distributions. Upper Panels and Lower Left
panel assume $R_{R} = 1.3$. 
{\it Upper Left:} $50\% $~of the galaxies artificially 
shifted to $\Delta_{a}R = 0\fm05$, and $50\% $ unaltered. 
{\it Upper Right:} $67 \% $~of the galaxies shifted to $\Delta_{a}R = 0\fm075$ 
and $33 \% $ unchanged.  
{\it Lower Left:} $80 \% $ at $\Delta_{a}R = 0\fm075$ and $20\% $ at 0.  
{\it Lower Right:} Here we set $R_{R} = 0.5$, with $33\%$ of the galaxies
shifted to $\Delta_{a}R = 0\fm025$, $33\%$ shifted to $0\fm05$, and $33\%$ 
remained at 0. The local minima in $\xi$ in each case correctly recover our 
input values.}
\label{bimodal}
\epsscale{1.0}
\end{figure}

To implement the third type of dust, which corresponds to dust distributed 
patchily, i.e. in `pockets' we randomly picked a specified
fraction of galaxies from the Cluster Group and assigned them exactly the 
same `dust' induced shifts in magnitude and color.
Other Cluster Group galaxies were assumed to lie behind cluster areas that
contained no dust.  The fraction of galaxies that were affected by
dust is the dust covering factor of the cluster. Figure~\ref{bimodal} shows 
examples of type (iii) dust. The upper right and lower left show results 
for bimodal dust: a fraction of the sky in Cluster Groups 
($33\%$ and $20\%$, respectively) is assumed to be devoid of dust and 
results in a local minimum in the 
$\xi_{k,l}$ plane at $(0,0)$. The rest of the sky area is obscured, and 
results in the second local minimum at non-zero values of $\Delta_a R$ and
$\Delta_a (B_J-R)$. The lower right panel 
shows the expected effect resulting from a cluster that contains
unobscured lines of sight as well as two different populations of dust,
both having $R_{R} = 0.5$ (instead of $1.3$ used in all other tests),
but different degrees of concentration.

Figure~\ref{bimodal} also tests the ability of our technique to detect dust 
distributed in `pockets' of small covering factors. It appears that dust with 
covering factors less than about $20\%$ would be difficult to detect, at least 
with the present sample of clusters.

Figures~\ref{dist} and \ref{bimodal} demonstrate that our technique can detect 
cluster dust, and distinguish various types of its spatial distribution:
smooth, non-uniform, or clumpy, as long as the typical values of 
%$\langle\Delta_a R\rangle$ and 
$\sigma_{a,R}$ are not too large,  $\,\lesssim 0\fm025-0\fm05$, and the 
covering factor is not too small,  $\simgt\, 20\%$.
It is important to note that these limitations as well as the CM-plane 
resolution are not limitations of the method, 
but are instead due to the limited number galaxies, and the relatively
bright flux limit of the APM galaxies. Our technique works well as long
as the dust induced shifts in color and magnitude are $\lesssim 1\%$
of the total CM-plane extent of the $n_{clus}$ and $n_{cont}$ distributions.
The extent of these in magnitude are limited on the faint side by the 
flux limit of the galaxy survey, while on the bright side the distributions
are effectively limited by the limited sky coverage.

\section{Summary and Conclusions}
\label{sc}

In this paper we have developed a technique to detect reddening and obscuration
due to widely-distributed dust in galaxy clusters. We have applied the method
to a sample of low redshift, $z\leq 0.08$ high Galactic Latitude APM clusters.
Our analysis indicates that the reddening and extinction due to dust within
$1.3$ Mpc of cluster centers relative to that in annuli 
immediately surrounding clusters are $|\Delta(B_{J} - R)| \leq 0\fm 025$ 
and $|\Delta R| \leq  0\fm 025 $, at $99\% $ confidence. Note that 
$|\Delta(B_{J} - R)| \leq 0\fm 025$ corresponds approximately to 
$|\Delta(B - V)| \leq 0\fm 020$.
Our result is
fully consistent with no dust being present on $\sim$ Mpc scales in moderately 
rich and poor clusters. 

There are several possible reasons for the non-detection of dust in our sample. 
First, it is possible that dust lying in galaxy clusters may in fact be 
clumped into self-shielded pockets with a small covering factor, $\lesssim 20\%$.
Our analysis would be relatively insensitive to this distribution
because only a minority of background galaxies would have been 
affected. It is also possible that dust is distributed over greater spatial 
regions then those we have tested in this work.  Because we are
somewhat limited by the size of the APM plates, we tested dust content at 
$< 1.3$ Mpc scales. Another possibility is that dust does not exist in 
significant quantities in the APM Galaxy Clusters.  
This stance is consistent with \citet{PTFV00}, who on the basis of 
estimates of dust insertion rates, anticipated an extinction of only 
$A_{B} = 0\fm 005 $ in the Virgo cluster, insomuch as the Virgo cluster 
is similar to APM clusters.  

The most likely interpretation of our results is that APM clusters do not
contain any dust. The spatial regions of moderately rich and poor APM 
clusters that we probe, $\simgt$ Mpc, 
roughly correspond to the same $\,R_{virial}$--normalized regions that
were probed by earlier studies of rich clusters and galaxy groups: 
$\sim$~few Mpc regions around rich clusters and $\lesssim$ Mpc regions 
around groups. These are beyond the virial radii of these systems, 
and probably correspond to regions where infall dominates. 
Combining our results with those of earlier studies, 
we conclude that rich, moderately rich, and poor clusters contain no dust 
in their infall regions, while galaxy groups do \citep{GMGM92}. 
The physical implications of this finding are beyond the scope 
of the present paper.

We tested our method by artificially reddening and obscuring galaxies 
belonging to the Cluster Groups. We applied various types of dust distribution
on the sky: uniform, non-uniform, and clumpy, and demonstrated that our 
technique correctly recovers the amplitude of reddening and obscuration, 
the value of total-to-selective extinction, and the approximate dust 
covering factor. The ability  of our method to recover these quantities
without relying on a priori assumptions about properties of dust makes
the method a powerful tool for detecting widely-distributed dust.

Our technique can be applied to any set of clusters or groups 
constructed from a homogeneous and uniform catalog of galaxies with at
least two colors. The performance of the method will improve if the galaxy 
distributions cover
a large range in magnitude and color. As we pointed out in
Section~\ref{tests}, dust induced shifts in color and magnitude are easier
to detect if they represent a small fraction of the total extent of the 
CM-plane distribution. 
Deeper photometric galaxy surveys would extend the faint magnitude end of
the distributions, while surveys with large sky coverage would
extend the bright end. Larger sky coverage would also enable the
use of smaller CM-plane pixels thus improving the resolution of the
method. Because of large plate-to-plate magnitude variation of the APM, 
our current analysis is restricted somewhat by the angular size 
of the plates. A large homogeneous survey would not have this restriction.  
Galaxy catalogs with band passes separated
to a larger degree than $B_J$ and $R$ would provide a longer lever arm
which would enable a better estimate of extinction.

\clearpage

\clearpage

\clearpage

\clearpage

\clearpage

\end{document}